\documentclass[aps,pre,twocolumn,nofootinbib,floatfix]{revtex4}
\usepackage{amsmath}
\usepackage{epsfig}

\input{tcilatex}

\begin{document}

\title{Resonant tunneling of quantum dot in a microcavity}
\author{Yueh-Nan Chen}
\email{ynchen.ep87g@nctu.edu.tw}
\author{Der-San Chuu}
\email{dschuu@cc.nctu.edu.tw}
\date{\today }

\begin{abstract}
We propose to measure Purcell effect by observing the current through a
semeiconductor quantum dot embedded in a microcavity. An electron and a hole
are injected separately into the quantum structure to form an exciton and
then recombine radiatively. The stationary current is shown to be altered if
one varies the cavity length or the exciton energy gap. Therefore, the
Purcell effect can be observed experimentally by measuring the current
through the quantum structure. In addition, we also find superradiance of
excitons between quantum dots may also be observed in an electrical way.

PACS: 78.67.-n, 42.50.Fx, 73.23.Hk
\end{abstract}

\maketitle

\affiliation{Department of Electrophysics, NationalW Chiao-Tung University, Hsinchu
30050, Taiwan}





Recently, much attention has been focused on enhanced spontaneous emission
(SE) rate of the quantum dot exciton in a optical microcavity. Historically,
the idea of controlling the SE rate by using a cavity \ was introduced by
Purcell\cite{1}. Considering the interaction between the atomic dipole and
the electromagnetic fields inside a cavity, the SE rate can be expressed as $%
(2\pi /\hbar )$ $\rho _{cav}(\omega )\left| \left\langle f\left| V\right|
i\right\rangle \right| ^{2}$ , where $\rho _{cav}(\omega )$ and$V$ are the
photon density of states and atom-vacuum field interaction hamiltonian,
respectively. For a planar cavity with distance $L_{c}$ between two mirrors,
the photon density of states is $N_{c}\omega /2\pi c^{2}$, where $N_{c}$ is
an integer less than $2L_{c}/\lambda $. Thus, by varying the cavity length $%
L_{c}$, the SE rate can be altered. The enhanced and inhibited SE rate for
the atomic system was intensively investigated in the 1980s\cite{2,3,4,5} by
using atoms passed through a cavity.

Turning to semiconductor systems, the electron-hole pair is naturally a
candidate for examining the spontaneous emission. However, as it was well
known, the excitons in a three dimensional system will couple with photons
to form polaritons--the eigenstate of the combined system consisting of the
crystal and the radiation field which does not decay radiatively\cite{6}.
Thus, in a bulk crystal, the exciton can only decay via impurity, phonon
scatterings, or boundary effects. The exciton can render radiative decay in
lower dimensional systems such as quantum wells, quantum wires, or quantum
dots as a result of broken symmetry. The decay rate of the exciton is
superradiant enhanced by a factor of $\lambda /d$ in a 1D system\cite{7} and
($\lambda /d)^{2}$ for 2D exciton-polariton\cite{8,9}, where $\lambda $ is
the wave length of emitted photon and $d$ is the lattice constant of the 1D
system or the thin film. In the past decades, the superradiance of excitons
in these quantum structures have been investigated intensively\cite%
{10,11,12,13}.

With the advances of modern fabrication technology, it has become possible
to fabricate the planar microcavities incorporating quantum wells\cite{14}
or quantum wires\cite{15}. In these systems it is possible to observe the
modified spontaneous emission rate of excitons. Similar to its decay-rate
counterpart, the frequency shift of a quantum wire exciton should also be
modified in a planar microcavity. By using the renormalization procedure
proposed by Lee \textit{et al}.\cite{16}, we have recently shown that the
frequency shift shows discontinuities at resonant modes\cite{17}. Instead of
one-dimensional confinement of photon fields, experimentalists are now able
to fabricate the quantum dot systems in laterally structured microcavities
that exhibit photon confinement in all three dimensions\cite{18,19}. Both
inhibition and enhancement of the spontaneous emission of quantum dot
excitons have been observed\cite{20}. However, acceptable experimental data
are still not plentiful owing to the difficulty of techniques in observing
the enhanced spontaneous emission optically. In this letter, a relatively
simple way to observe the enhanced spontaneous emission is proposed to embed
the quantum ring or the quantum dot in a microcavity\cite{21}. By injecting
electron and hole into the quantum dot, a photon is generated by the
recombination of the exciton. This process allows one to determine Purcell
effect by measuring the current through the quantum dot.

In our model, we consider a quantum dot embedded in a \textit{p-i-n}
junction, which is similar to the device proposed by O. Benson \textit{et al}%
\cite{21}. The energy-band diagram is shown in Fig.1.

\begin{figure}[th]
\includegraphics[width=7.5cm]{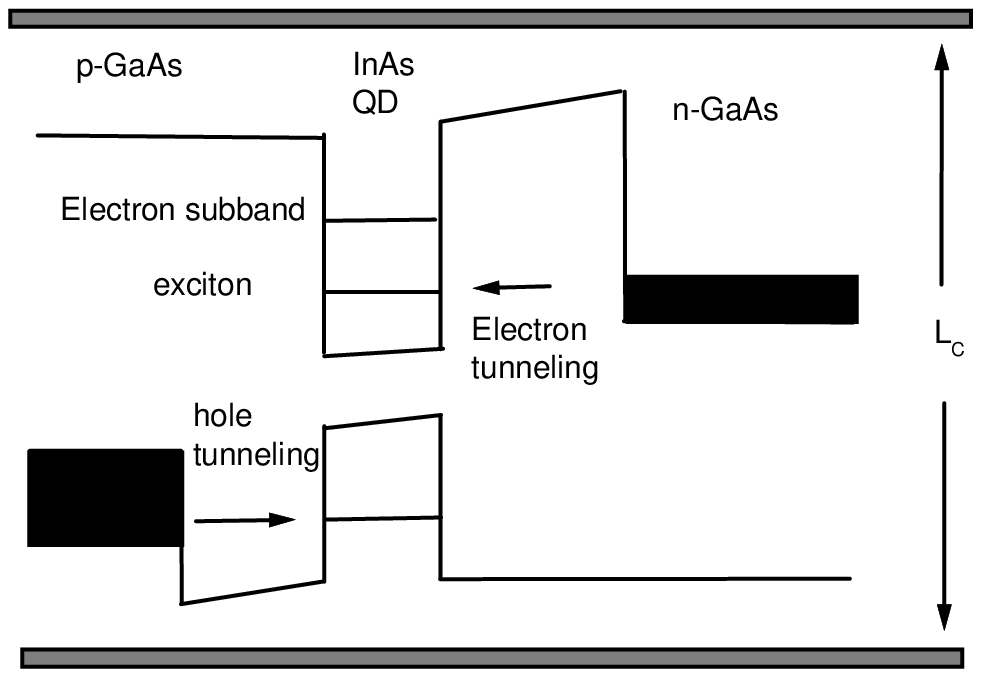}
\caption{Energy-band diagram of the structure.}
\end{figure}

Both the hole and electron reservoirs are assumed to be in thermal
equilibrium. For the physical phenomena we are interested in, the fermi
level of the \textit{p(n)}-side hole (electron) is slightly lower (higher)
than the hole (electron) subband in the dot. After a hole is injected into
the hole subband in the quantum dot, the \textit{n}-side electron can tunnel
into the exciton level because of the Coulomb interaction between the
electron and hole. Thus, we may assume three dot states

\begin{eqnarray}
\left| 0\right\rangle &=&\left| 0,h\right\rangle  \notag \\
\left| U\right\rangle &=&\left| e,h\right\rangle  \notag \\
\text{ }\left| D\right\rangle &=&\left| 0,0\right\rangle
\end{eqnarray}%
,where $\left| 0,h\right\rangle $ means there is one hole in the quantum
dot,\ $\left| e,h\right\rangle $ is the exciton state, and $\left|
0,0\right\rangle $ represents the ground state with no hole and electron in
the quantum dot. One might argue that one can not neglect the state $\left|
e,0\right\rangle $ for real device since the tunable variable is the applied
voltage. This can be resolved by fabricating a thicker barrier on the
electron side so that there is little chance for an electron to tunnel in
advance. We can now define the dot-operators $\overset{\wedge }{n_{U}}\equiv
\left| U\right\rangle \left\langle U\right| ,$ $\overset{\wedge }{n_{D}}%
\equiv \left| D\right\rangle \left\langle D\right| ,$ $\overset{\wedge }{p}%
\equiv \left| U\right\rangle \left\langle D\right| ,$ $\overset{\wedge }{%
s_{U}}\equiv \left| 0\right\rangle \left\langle U\right| ,$ $\overset{\wedge 
}{s_{D}}\equiv \left| 0\right\rangle \left\langle D\right| $. The total
hamiltonian $H$ of the system consists of three parts: the dot hamiltonian,
the photon bath, and the electron (hole) reservoirs:

\begin{eqnarray}
H &=&H_{0}+H_{T}+H_{V}  \notag \\
H_{0} &=&\varepsilon _{U}\overset{\wedge }{n_{U}}+\varepsilon _{D}\overset{%
\wedge }{n_{D}}+H_{p}+H_{res}  \notag \\
H_{T} &=&\sum_{k}g(D_{k}b_{k}^{\dagger }\overset{\wedge }{p}+D_{k}^{\ast
}b_{k}\overset{\wedge }{p}^{\dagger })=g(\overset{\wedge }{p}X+\overset{%
\wedge }{p}^{\dagger }X^{\dagger })  \notag \\
H_{p} &=&\sum_{k}\omega _{k}b_{k}^{\dagger }b_{k}  \notag \\
H_{V} &=&\sum_{\mathbf{q}}(V_{\mathbf{q}}c_{\mathbf{q}}^{\dagger }\overset{%
\wedge }{s_{U}}+W_{\mathbf{q}}d_{\mathbf{q}}^{\dagger }\overset{\wedge }{%
s_{D}}+c.c.)  \notag \\
H_{res} &=&\sum_{\mathbf{q}}\varepsilon _{\mathbf{q}}^{U}c_{\mathbf{q}%
}^{\dagger }c_{\mathbf{q}}+\sum_{\mathbf{q}}\varepsilon _{\mathbf{q}}^{D}d_{%
\mathbf{q}}^{\dagger }d_{\mathbf{q}}.
\end{eqnarray}%
In above equation, $b_{k}$ is the photon operator, $gD_{k}$ is the dipole
coupling strength, $X=\sum_{k}D_{k}b_{k}^{\dagger }$ ,$\ $and $c_{\mathbf{q}%
} $ and $d_{\mathbf{q}}$ denote the electron operators in the left ad right
reservoirs, respectively. Here, $g$ is a constant with a unit of the
tunneling rate. The couplings to the electron and hole reservoirs are given
by the standard tunnel hamiltonian $H_{V},$ where $V_{\mathbf{q}}$ and $W_{%
\mathbf{q}}$ couple the channels $\mathbf{q}$ of the electron and the hole
reservoirs. If the couplings to the electron and the hole reservoirs are
weak, then it is reasonable to assume that the standard Born-Markov
approximation with respect to these couplings is valid. In this case, one
can derive a master equation from the exact time-evolution of the system.
The equations of motion can be expressed as (cp. [22])

\begin{eqnarray*}
\overset{\wedge }{\left\langle n_{U}\right\rangle }_{t}-\overset{\wedge }{%
\left\langle n_{U}\right\rangle }_{0} &=&-ig\int_{0}^{t}dt^{\prime }\{%
\overset{\wedge }{\left\langle p\right\rangle }_{t^{\prime }}-\overset{%
\wedge }{\left\langle p^{\dagger }\right\rangle }_{t^{\prime }}\} \\
&&+2\Gamma _{U}\int_{0}^{t}dt^{\prime }(1-\overset{\wedge }{\left\langle
n_{U}\right\rangle }_{t^{\prime }}-\overset{\wedge }{\left\langle
n_{D}\right\rangle }_{t^{\prime }})
\end{eqnarray*}

\begin{equation*}
\overset{\wedge }{\left\langle n_{D}\right\rangle }_{t}-\overset{\wedge }{%
\left\langle n_{D}\right\rangle }_{0}=-ig\int_{0}^{t}dt^{\prime }\{\overset{%
\wedge }{\left\langle p\right\rangle }_{t^{\prime }}-\overset{\wedge }{%
\left\langle p^{\dagger }\right\rangle }_{t^{\prime }}\}-2\Gamma
_{D}\int_{0}^{t}dt^{\prime }\overset{\wedge }{\left\langle
n_{D}\right\rangle }_{t^{\prime }}
\end{equation*}

\begin{eqnarray*}
\overset{\wedge }{\left\langle p\right\rangle }_{t}-\overset{\wedge }{%
\left\langle p\right\rangle _{t}^{0}} &=&-\Gamma _{D}\int_{0}^{t}dt^{\prime
}e^{i\varepsilon (t-t^{\prime })}\left\langle X_{t}X_{t^{\prime }}^{\dagger }%
\widetilde{p}(t^{\prime })\right\rangle _{t^{\prime }} \\
&&-ig\int_{0}^{t}dt^{\prime }e^{i\varepsilon (t-t^{\prime })}\{\left\langle 
\overset{\wedge }{n_{U}}X_{t}X_{t^{\prime }}^{\dagger }\right\rangle
_{t^{\prime }} \\
&&-\left\langle \overset{\wedge }{n_{D}}X_{t^{\prime }}^{\dagger
}X_{t}\right\rangle _{t^{\prime }}\}
\end{eqnarray*}

\begin{eqnarray}
\overset{\wedge }{\left\langle p^{\dagger }\right\rangle }_{t}-\overset{%
\wedge }{\left\langle p\right\rangle _{t}^{0}} &=&-\Gamma
_{D}\int_{0}^{t}dt^{\prime }e^{-i\varepsilon (t-t^{\prime })}\left\langle 
\widetilde{p}^{\dagger }(t^{\prime })X_{t^{\prime }}X_{t}^{\dagger
}\right\rangle _{t^{\prime }}  \notag \\
&&+ig\int_{0}^{t}dt^{\prime }e^{-i\varepsilon (t-t^{\prime })}\{\left\langle 
\overset{\wedge }{n_{U}}X_{t^{\prime }}X_{t}^{\dagger }\right\rangle
_{t^{\prime }}  \notag \\
&&-\left\langle \overset{\wedge }{n_{D}}X_{t}^{\dagger }X_{t^{\prime
}}\right\rangle _{t^{\prime }}\}.
\end{eqnarray}%
, where $\Gamma _{U}$ $=2\pi \sum_{\mathbf{q}}V_{\mathbf{q}}^{2}\delta
(\varepsilon _{U}-\varepsilon _{\mathbf{q}}^{U})$ , $\Gamma _{D}=2\pi \sum_{%
\mathbf{q}}W_{\mathbf{q}}^{2}\delta (\varepsilon _{D}-\varepsilon _{\mathbf{q%
}}^{D})$, and $\varepsilon =\varepsilon _{U}-\varepsilon _{D}$ is the energy
gap of the quantum dot exciton. Here, $\widetilde{p}(t^{\prime
})=pe^{i\varepsilon t}X_{t^{\prime }}$, and $X_{t^{\prime }}$ denotes the
time evolution of $X$ with $H_{p}$. The expectation value $\overset{\wedge }{%
\left\langle p^{(\dagger )}\right\rangle _{t}^{0}}$describes the decay of an
initial polarization of the system and plays no role for the stationary
current. Therefore, we shall assume the initial expectation value of $%
\overset{\wedge }{p}^{(\dagger )}$ vanishes at time $t=0$.

As can be seen from Eq. (3), there are terms like $\left\langle \overset{%
\wedge }{n_{U}}X_{t}X_{t^{\prime }}^{\dagger }\right\rangle _{t^{\prime }}$
which contain products of dot operators and photon operators. If we are
interested in small coupling parameters here, a decoupling of the reduced
density matrix $\widetilde{\rho }(t^{\prime })$ can be written as

\begin{equation}
\widetilde{\rho }(t^{\prime })\approx \rho _{ph}^{0}Tr_{ph}\widetilde{\rho }%
(t^{\prime }).
\end{equation}
By using the above equation, we obtain

\begin{equation}
Tr(\widetilde{\rho }(t^{\prime })\overset{\wedge }{n_{U}}X_{t}X_{t^{\prime
}}^{\dagger })\approx \overset{\wedge }{\left\langle n_{U}\right\rangle }%
_{t^{\prime }}\left\langle X_{t}X_{t^{\prime }}^{\dagger }\right\rangle _{0}
\end{equation}%
and correspondingly the other products of operators can be obtained also.
For spontaneous emission, the photon bath is assumed to be in equilibrium.
The expectation value $\left\langle X_{t}X_{t^{\prime }}^{\dagger
}\right\rangle _{0}\equiv C(t-t^{\prime })$ is a function of the time
interval only. We can now define the Laplace transformation for real $z,$

\begin{eqnarray}
\overset{}{C_{\varepsilon }}(z) &\equiv &\int_{0}^{\infty
}dte^{-zt}e^{i\varepsilon t}C(t)  \notag \\
\overset{}{n_{U}}(z) &\equiv &\int_{0}^{\infty }dte^{-zt}\overset{\wedge }{%
\left\langle n_{U}\right\rangle }_{t}\text{ \ }etc.,\text{ }z>0
\end{eqnarray}%
and transform the whole equations of motion into $z$-space,

\begin{eqnarray}
n_{U}(z) &=&-i\frac{g}{z}(p(z)-p^{\ast }(z))+2\frac{\Gamma _{U}}{z}%
(1/z-n_{U}(z)-n_{D}(z))  \notag \\
n_{D}(z) &=&\frac{g}{z}(p(z)-p^{\ast }(z))-2\frac{\Gamma _{D}}{z}n_{D}(z) \\
p(z) &=&-ig\{n_{U}(z)C_{\varepsilon }(z)-n_{D}(z)C_{-\varepsilon }^{\ast
}(z)\}-\Gamma _{D}p(z)C_{\varepsilon }(z)  \notag \\
p^{\ast }(z) &=&ig\{n_{U}(z)C_{\varepsilon }^{\ast
}(z)-n_{D}(z)C_{-\varepsilon }(z)\}-\Gamma _{D}p^{\ast }(z)C_{\varepsilon
}^{\ast }(z).  \notag
\end{eqnarray}%
These equations can then be solved algebraically. The tunnel current $%
\widehat{I}$ can be defined as the change of the occupation of $\overset{%
\wedge }{n_{U}}$ and is given by $\widehat{I}\equiv ig(\overset{\wedge }{p}-%
\overset{\wedge }{p}^{\dagger }),$ where we have set the electron charge $%
e=1 $ for convenience. The time dependence of the expectation value $\overset%
{\wedge }{\left\langle I\right\rangle }_{t}$ can be obtained by solving Eq.
(7) and performing the inverse Laplace transformation. For time $%
t\rightarrow \infty ,$ the result is

\begin{eqnarray}
\overset{\wedge }{\left\langle I\right\rangle }_{t\rightarrow \infty } &=&%
\frac{2g^{2}\Gamma _{U}\Gamma _{D}B}{g^{2}\Gamma _{D}B+[g^{2}B+\Gamma
_{D}+2\gamma \Gamma _{D}^{2}+(\gamma ^{2}+\Omega ^{2})\Gamma _{D}^{3}]} 
\notag \\
B &=&\gamma +(\gamma ^{2}+\Omega ^{2})\Gamma _{D},
\end{eqnarray}%
where $g^{2}\Omega $ and $g^{2}\gamma $ are the exciton frequency shift and
decay rate, respectively. The derivation of the current equation is closely
analogous to the spontaneous emission of phonons in double dots\cite{22}, in
which the correlation functions $\left\langle X_{t}X_{t^{\prime }}^{\dagger
}\right\rangle _{0}$ is given by the electron-phonon interaction.

Since the stationary current through the quantum dot depends strongly on the
decay rate $\gamma $, the results of a quantum dot inside a planar
microcavity is numerically displayed in Fig. 2. In plotting the figure, the
current is in terms of 100 pA, and the cavity length is in units of $\lambda
_{0}/2$, where $\lambda _{0}$ is the wavelength of the emitted photon.
Furthermore, the tunneling rates, $\Gamma _{U}$ and $\Gamma _{D}$, are
assumed to be equal to 0.2$\gamma _{0}$ and $\gamma _{0},$ respectively.
Here, a value of 1/1.3ns for the free-space quantum dot decay rate $\gamma
_{0}$ is used in our calculations\cite{19}. Also, the planar microcavity has
a Lorentzian broadening at each resonant modes (with broadening width equals
to 1\% of each resonant mode)\cite{17}. As the cavity length is less than
half of the wavelength of the emitted photon, the stationary current is
inhibited. This is because the energy of the photon generated by the quantum
dot is less than the cut-off frequency of the planar microcavity. Moreover,
the current is increased whenever the cavity length is equal to multiple
half wavelength of the emitted photon. It represents as the cavity length
exceeds some multiple wavelength, it opens up another decay channel abruptly
for the quantum dot exciton, and turns out that the current is increased.
With the increasing of cavity length, the stationary current becomes less
affected by the cavity and gradually approaches to free space limit.

\begin{figure}[th]
\includegraphics[width=8cm,clip=true]{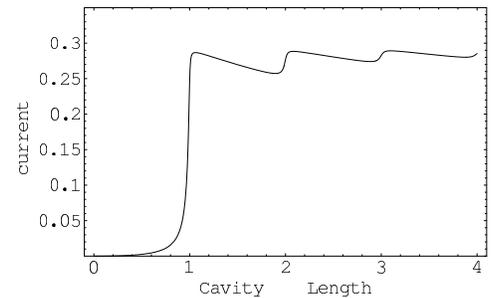}
\caption{Stationary tunnel current, Eq. (8), as a function of cavity length $%
L_{c}$. The vertical and horizontal units are 100 pA and $\protect\lambda %
_{0}$, respectively.}
\end{figure}

To understand the inhibited current thoroughly, we now fix the cavity length
equal to $\lambda _{0}/2$ and vary the exciton energy gap, while the planar
microcavity is now assumed to be perfect. The vertical and horizontal units
in Fig. 3 are 100 pA and $2hc/\lambda _{0}$, respectively. Here, $\lambda
_{0}$ is the wavelength of the photon emitted by the quantum dot exciton in
free space. Once again, we observe the suppressed current as the exciton
energy gap is tuned below the cut-off frequency. The plateau features in
Fig. 3 also comes from the abruptly opened decay channels for the quantum
dot exciton. From the experimental point of view, it is not possible to tune
either the cavity length or the energy gap for such a wide range. A possible
way is to vary the exciton gap around the first discontinuous point $%
2hc/\lambda _{0}$. Since the discontinuities should smear out for the real
microcavity, it is likely to have a peak if one measures the differential
conductance $d\overset{\wedge }{\left\langle I\right\rangle }/d\varepsilon $
as a function of energy gap $\varepsilon $.

\begin{figure}[th]
\includegraphics[width=8cm]{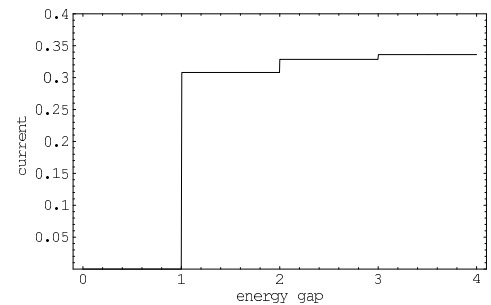}
\caption{Stationary current as a function of exciton energy gap $\protect%
\varepsilon $. The cavity length is fixed to $\protect\lambda _{0}/2$. The
current is in units of 100 pA, while the energy gap is terms of $2hc/\protect%
\lambda _{0}$.}
\end{figure}

Our proposal can also be used to measure the superradiance of quantum dots
in an electrical way. Consider now the system containing two quantum dots
with a distance $d$ as shown in Fig. 4(a). One of the obstacles in measuring
the superradiance between the quantum dots comes from the random size of the
dots which result in a random distribution of energy gap. This can be
overcome by growing two gates above the quantum dots. The energy gap and the
orientation of the dipole moments can be controlled well. Analogous to the
two-ion system\cite{23}, the electron and hole can tunnel into the
superradiant or subradiant state. The corresponding decay rate for the two
channels is given by

\begin{equation}
\gamma _{\pm }=\gamma _{0}(1\pm \frac{\sin (2\pi d/\lambda _{0})}{2\pi
d/\lambda _{0}}),
\end{equation}%
where the two signs $\pm $ correspond to the two different relative
orientations of the dipole moments of the two dots. Fig. 4(b) shows the
stationary currents of the superradiant (solid line) and subradiant (dashed
line) channels. The interference effect between the dots is displayed
explicitly. In principle, one can incorporate more quantum dots in the
system, and many superradiant effects can be examined by the electrical
current.

\begin{figure}[th]
\includegraphics[width=8.2cm,clip=true]{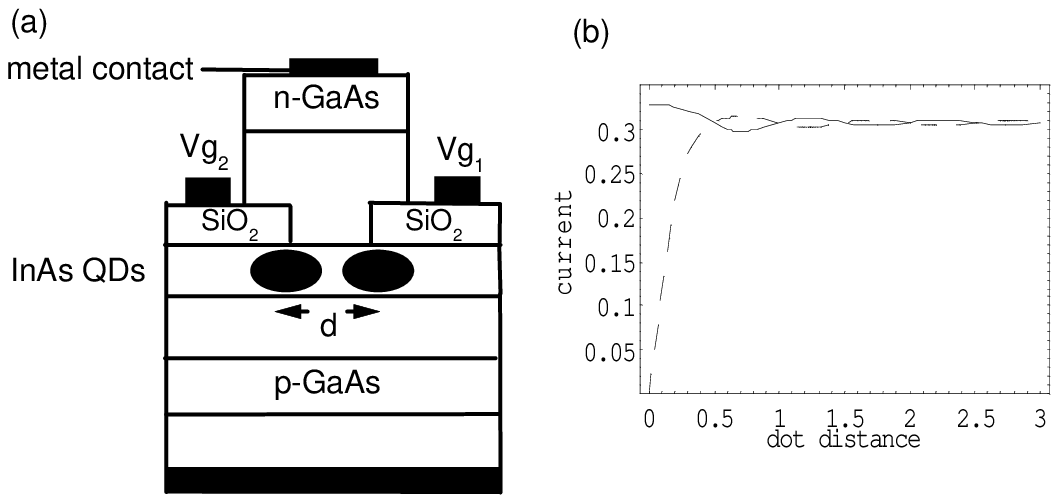}
\caption{(a) Proposed device structure. Two InAs quantum dots are embedded
in a\textit{\ p-i-n }junction. Above the quantum dots are two gates, which
control the energy gap and orientation of the diploes. (b) Stationary
current (in units of 100 pA) through the superradiant (solid channel) and
subradiant (dashed line) channel as a function of dot distance $d$ (in units
of $\protect\lambda _{0}$).}
\end{figure}

In conclusion, we have proposed a method of detecting the Purcell effect in
a semiconductor quantum dot system. By incorporating the InAs quantum dot
between a \textit{p-i-n} junction surrounded by a planar microcavity, the
Purcell effect on stationary tunnel current can be examined either by
changing the cavity length or by varying the exciton energy gap. Second, it
is also possible to observe the superradiant effects between two dots by
using present model. The interference features are pointed out and may be
observable in a suitably designed experiment.

We would like to thank to Prof. D. A. Rudman and Prof. M. Keller of NIST for
helpful discussions. One of authors (Y. N. Chen) also appreciate valuable
discussions with P. C. Chen of UCSD. This work is supported partially by the
National Science Council, Taiwan under the grant number NSC
90-2112-M-009-026.

\end{document}